\documentclass[5p]{elsarticle}

\usepackage{fancyvrb}
\usepackage{hyperref}
\usepackage{longtable}

\journal{Astronomy and Computing, elsevier}

\begin{document}

\begin{frontmatter}

\title{KERN}

\author[ska,rhodes]{Gijs Molenaar}
\author[rhodes,ska]{Oleg Smirnov}

\address[ska]{SKA Sounth Africa, 3rd Floor, The Park, Park Road, Pinelands, 7405, South Africa}
\address[rhodes]{Department of Physics \& Electronics, Rhodes University, PO Box 94, Grahamstown, 6140, South Africa}

\date{\today}

\begin{abstract}
KERN is a bi-annually released set of radio astronomical software packages. It should contain most of the 
standard tools that a radio astronomer needs to work with radio telescope data.
The goal of KERN is to save time and prevent frustration in setting up of scientific pipelines, and to assist
in achieving scientific reproducibility.
\end{abstract}


\end{frontmatter}

\section{Introduction}
The installation of scientific software for use in astronomy can be notoriously challenging. The radio astronomy
community has a limited number of dedicated software engineers and often lacks the human resources to dedicate to
industrial-quality software development and maintenance. It is not uncommon that poorly written and badly maintained
software packages of high complexity are used by scientists around the world, as these provide some set of algorithmic
features not available elsewhere.

KERN has been created to facilitate scientists and system administrators. KERN is the name
of the project to structure and automate the packaging of scientific software. The main deliverable
is the KERN suite, a bi-annually released set of 3rd party open source scientific software packages.

The primary goals of KERN are: to make it easier to install the scientific software, to supply a
consistent working environment to a scientist and to improve interoperability and interchangeability.

Due to human resource limitations, we target KERN to one operating system and distribution. This is
unfortunate, but recent development and adaptation of containerisation technology make it
easier to deploy packaged software on most platforms. Limiting us to only one platform enables us to
focus on performing the packaging only once, and to do this well. The choice of this one platform is
then based on install base (desktop, server) and ease of use for user and developer (package creator).

The intended audience of this paper is threefold:
\begin{itemize}
\item the user, who is interested in using the software bundled with KERN,
\item the developer, who wants his radio astronomy software available to a wider range of users and
\item the system administrator who is setting up systems intended to be used for radio astronomical data reduction.
\end{itemize}

The name KERN means `core' in Dutch and Afrikaans.

\section{The target platform}

A quick look around various astronomy institutes and universities
shows that GNU/Linux and OS X are the most used personal computing platforms. On the server side it is
without question GNU/Linux. Compared to OS X, GNU/Linux is an open source and a freely available
platform, which is also a clear advantage. These facts combined result in the choice for Linux as the
KERN platform.

However, further consideration was required before selection of the most suitable platform. There are various
flavours of GNU/Linux, with different design philosophies and varying packaging formats. The most popular
distributions can be split into two groups, RPM (Red Hat Package Manager) package and Debian package based
distributions. There is no major advantage or disadvantage to either package format. Although there are diverse local
trends, it is our experience that in the South African radio astronomy community, the majority of frequently utilised
platforms are Debian based, specifically Ubuntu LTS. This distribution also appears to have popularity worldwide.
Therefore, it was the most logical choice as KERNs target platform.

\section{Other packaging methods}

In this section we discuss other packaging systems available to us, and motivate our choice
not to use these.

\subsection{Anaconda}
A packaging effort named Anaconda is currently gaining popularity. Anaconda is a cross
platform set of scientific software, with a focus on Python. It supports GNU/Linux, Windows and OS X.
OS X is also often used as a desktop environment in radio astronomy. Supporting OS X would be advantageous for
many end users.

We have performed experiments with packaging packages for Anaconda. Users have reported that Anaconda is easy to
operate; however, at the time of writing, the packaging procedure is cumbersome for the developer. In effect,
developers cannot generate the same high quality, seamlessly installable packages, as is achievable with native Linux
packaging methods. Additionally, Anaconda lacks an equivalent to Debians Lintian, a packaging tool that dissects
a Debian package and tries to find bugs and violations of the Debian policies.

Various software packages in KERN are not created with OS X support in mind thus requiring various modifications to
the source code. Also, compilation procedures can vary greatly across platforms, doubling the packaging and maintenance
effort if we would support OS X as a platform.

In addition, Linux distributions come with a large set of prepackaged software, which eliminates
the need to package many dependencies. Using Anaconda would necessitate packaging up many dependencies ourselves.

The limitations of Anaconda led to the preference of Debian- over Anaconda packages.

\subsection{Python and pip}
The Python programming language has become the most widely used language in astronomy \cite{Momcheva2015Software}.
Python comes with a package manager called pip. Pip assists
in downloading and installing Python packages from the Python package index (PyPi). Another useful
tool for setting up Python environments is called {\tt virtualenv}. Virtualenv enables a user to set up
one or more isolated Python environments without system administrator rights. The combination of these two tools
 enables the end user to set up various custom environments with specific versions of
dependencies. For pure Python projects, pip and virtualenv are cross platform and independent of the
host operating system package manager. However, pip is less suited for impure Python projects. Some Python
libraries depend on non-Python run time libraries and/or non-Python development headers compile
time, making them ``impure''. A recent improvement to the Python Packaging system is the introduction of
wheels. Wheels are pre-compiled binary Python packages. These do not require compilation and will
work if the packaged library does not have unusual requirements. An example of a independent binary
wheel is Numpy. Numpy only depends on Python and a small set of system libraries. The Application
Binary Interface (ABI) differs across host platforms and python versions, requiring a wheel for every platform and
python combination. These are supplied on PyPi and the correct version is automatically selected by pip on installation.

Problems arise for example, with the python-casacore package, which has more unusual dependencies.
Python-casacore depends on the casacore package and both packages need
to have a matching ABI. If the version of casacore differs between compile time and
run time, the library does not work. Pip does not have any control over reinforcing the shared library
version. This would imply that we need to create, upload and maintain wheels for every casacore released.
Moreover, with every casacore release we would also need to create a wheel for every OS and supported
profile. The exponential growth of this cartesian product quickly becomes cumbersome for the package
maintainers.

These limitations combined with pip's inability to handle non-Python libraries, make pip an ill
suited candidate for our packaging effort. Nevertheless, this does not mean pip and KERN cannot be combined.

The approach we adopt is that we prepackage impure python libraries and bundle them with KERN. These
Python packages are then precompiled against a set of specific library versions. This guarantees the
ABIs always match up. Users can then augment the system python installation, or a Python virtual
environment, with packages from the packaging index using pip.

Although KERN supports both Python 2 and Python 3, most Python packages in KERN only support Python version 2. For
now, the only package that supports Python 3 is python-casacore. The KERN Python 3 package for casacore is named
python3-casacore.

\subsection{Collaboration with Debian}
Ubuntu is directly based on Debian and thus is similar to Debian. Nonetheless, due to version differences in the
bundled libraries in each distribution, KERN packages are unlikely to run on Debian. Fortunately the packaging procedure
is identical for Debian and Ubuntu, which makes creating true Debian packages a matter of a recompilation of the source
package.

In contrast, the build system, dependency management and library management for RPM is completely different. Porting
our packages to RPM is non-trivial, and maintaining support for RPM based distributions would imply doubling the required effort.

We have established a collaboration with Debian developers, and some packages from KERN (e.g. {\tt casacore}
and {\tt aoflagger}) have been incorporated into Debian directly. These packages have been uploaded to the Debian
archive, and changes are synchronised between KERN and the Debian archive.

Not all KERN packages are suitable for uploading to Debian. Packages with a small user base or packages that
are fragile and receive continuous fixes (as opposed to formal release) are not well suited to this distribution
model, since it can take some time before a package ends up in a Debian release. For the more stable packages in
KERN, we expect a continuation of this effort, with more packages ending up in Debian in the future.

\section{Usage}

To use the packages of KERN, one needs to add the KERN remote repository to the system. It is recommended to use the
latest released version, which is KERN-2 at time of writing. KERN-2 is packaged for Ubuntu 16.04: using the packages
on a different distribution or version will most likely fail. If running Ubuntu 16.04 is not an option on a
particular system, we recommend using Docker, Singularity (see below) or a virtual machine.

The {\tt add-apt-repository} command should be used to add the KERN repository to a system. Some packages in KERN
depend on Ubuntu packages in the {\tt multiverse}  and {\tt restricted} repositories (CUDA is an example of such a
dependency). Once the local cache is updated using {\tt apt-get update} the package cache can be searched using
{\tt apt-cache search {\em PACKAGE}} and packages can be installed using the {\tt apt-get install {\em PACKAGE}} command.

In case of an unexpected fault, it is important to ensure that the latest versions of all packages are being used
(by running {\tt apt update} and {\tt apt upgrade}), before reporting new issues.

Missing packages can be nominated for inclusion in KERN by requesting the packaging on the issue
tracker\footnote{\url{https://github.com/kernsuite/packaging/issues}}.

For further instruction on how to install and use KERN packages, we refer to the KERN suite
documentation\footnote{\url{http://kernsuite.info/}}.

\section{Notable packages}

Here we discuss the important or non-trivial packages part of KERN. Note that this is not the full list of packages. The
full list of packages contained in KERN-2 is provided in Table \ref{tab:packagelist}.

\subsection{casacore}
Most of the packages in KERN depend on casacore\cite{vanDiepen2015174}. Casacore is a suite of C++ libraries for radio
astronomy data processing. The most important library is the table system for working with Measurement Set, which is
currently the most used data storage format in radio astronomy. Since it is such an important package, additional
effort has been put in making this package quality high, and it should be seen as an example and reference for all
other packages. The package has also been accepted in the main Debian repository and updates are synchronised
between Debian and KERN\footnote{\url{https://packages.qa.debian.org/c/casacore.html}}.

\subsection{casacore data}
Casacore has a ``soft'' dependency on the ``casacore data'' package. The latter containers ephemerides, geodetic data
and other tables required for performing calculations such as coordinate conversions. Strictly speaking, the casacore
data package is not required if one is not calculating e.g. coordinate conversions, but in practice many components
of casacore will fail or give warnings if the data package is missing. The casacore data package
is updated on a regular basis using a cron job\footnote{\url{https://github.com/casacore/casacore-data-update}}.
Updated content typically consists of accurate GPS movements of the tectonic plates, new or updated radio telescope
positions, leap seconds, etc. There is no central authority controlling the content of the casacore data package, 
and various institutes around the world create their own versions. We base the KERN package on the data supplied 
and updated weekly by the Netherlands Institute for Radio Astronomy (ASTRON), which is published on its public FTP 
server\footnote{\url{ftp://ftp.astron.nl/outgoing/Measures/}}.

\subsection{MeqTrees}
MeqTrees\cite{noordam2010meqtrees} is a software suite developed at ASTRON and subsequently Rhodes and SKA SA.
MeqTrees is aimed at implementing various versions of the Radio Interferometry Measurement Equation
(RIME)\cite{smirnov2011rime1}, which is used for simulation and calibration of radio interferometric data. MeqTrees
consists of two core packages: Timba (the C++-based computational implementation) and Cattery (a set of
Python frameworks providing end-user tools). Other packages in the suite include Tigger (a sky model and FITS image
viewer), Owlcat (a set of Measurement Set manipulation utilities) and Pyxis (a scripting/pipelining framework).

\subsection{CASA}
CASA, Common Astronomy Software Applications\cite{2007McMullin}, is a popular system for
reduction of radio astronomy data. CASA is maintained by the National Radio Astronomy 
Observatory\footnote{\url{http://casa.nrao.edu/}}. It has proven to be one of the most challenging pieces of
software to package up. CASA is delivered as one monolithic self-contained tarball, which bundles a complete set
of dependencies, including its own python interpreter and IPython interface. This scheme has the advantage that it
runs on most systems without modification or additional library requirements.
However, the tarball is sizable (over 1Gb at time of writing). In addition, various
dependencies bundled with CASA are dated (for example, the bundled IPython package
is version 0.10, dating from 2010). This makes it hard to install updated packages into the CASA bundled 
Python, since CASA itself may break if packages are updated. Since CASA depends on so many old libraries, 
it is close to impossible to install it as an overlay on
the default Debian filesystem and make use of the system Python interpreter and libraries. Some effort has been made
towards unbundling an earlier version of CASA (4.3) into individual packages, but the CASA team was unable to dedicate
resources to this effort in further releases. We have therefore adopted the single large package option as the only
practical possibility for now. Other attempts have been made to bridge the gap between the system and CASA Python
interpreter, but the result is still far from ideal\cite{2015Staley}.

We have decided to take a pragmatic approach  and package a subset of CASA and instead of installing it as an overlay on
the system, choosing to do so in a separate directory . Having a package simplifies the installation and helps us to manage
dependencies, since there are packages that depend on CASA. We reduced the install size by unbundling the
experimental Carta viewer, which reduced the final install footprint by about 1Gb. In case of CASA 4.7.1 the tarball
is reduced by 60\% to about 400MB. The software is installed in {\tt /opt/casa} and a symlink to the casa binary is
created under {\tt /usr/bin}. The package works exactly like the naive CASA installation (apart from the omitted
Carta), but unfortunately still cannot interoperate with the Debian system Python installation. Since the package is
a modified version  of the original release, we decided to rename the package to {\tt casalite}.

\subsection{AIPS}
NRAO's venerable Astronomical Image Processing System (AIPS) is the predecessor of CASA and, despite its age,
maintains a wide and enthusiastic user base. AIPS has some of the oldest code of all the packages in KERN and
depends on various (nowadays rather arcane and archaic) system configuration modifications. We bundle
AIPS with a minimal configuration that assumes a single system (localhost) setup. All software is installed under
{\tt /opt/aips}, similar to CASA. The software requires a writable data folder in order to operate, thus
when the package is installed, an {\tt aips} group is created if it does not exist. Users who
want to use AIPS need to be added to this group. Due to the complexity of AIPS compilation, the KERN packages are 
not compiled from source, but are rather based on the binary tarball distribution.

\subsection{LOFAR}
Another notable package is the {\tt lofar} package. This package contains all the code bundled in the LOFAR imaging
pipeline. Most of it is used by astronomers working with data from the LOFAR radio telescope, however, it has a number
of general-interest components, such as the PyBDSM / PyBDSF source finder and the {\tt makems} Measurement Set
creation tool. For the reader familiar with building and using the LOFAR pipeline, it is relevant to know
that the full ``Offline'' bundle is contained in the KERN package.

\subsection{Pulsar software}

KERN contains a number of software packages from the pulsar community such as {\tt tempo}, {\tt presto} and
{\tt psrchive}. These tools were particularly difficult to create packages from, since most of them do not perform
normal release management and tend to have broken build scripts. For packages without version numbers, we introduce our
own date-based versioning scheme and the packages are updated on request.

\section{Containerisation}

\subsection{Docker}
Docker is a mature and popular concept for distributing software and managing processes.
Although it is easy to create a docker container from a piece of software, it is no replacement
for proper software packaging. We think that proper packaging and containerisation go hand in hand:
Debian packages provide robustness and dependency management while containerisation provides portability and
distributability.

We have prepared an easy to use base docker image which can be used to create custom docker images
containing all the KERN packages combined with end-user scripts. The Dockerfile below is all that is needed 
to set up a Docker container for any given package in KERN. The example below is for aoflagger:

\begin{Verbatim}[frame=single]
FROM kernsuite/base:2
RUN docker-apt-install aoflagger
\end{Verbatim}

The kernsuite Docker image is a clean Ubuntu system with the KERN suite repository enabled. It also
contains an up-to-date pip so that one can directly install Python libraries. The {\tt docker-apt-install}
command is just a wrapper script that updates the apt cache before installing the package, then removes the cache. The
latter is done to prevent cluttering of the Docker image, which could otherwise  lead to exploding image sizes.

\subsection{Singularity}

A number of assumptions made by Docker creators have created security concerns and have made Docker a poor fit to
the typical HPC environment. This has motivated the development of an alternative containerisation technology called
Singularity. While this is presently less popular, Singularity is more suited for deploying software in a
multi-tenant cluster environment, which is the most common environment in science. We have created scripts to 
easily set up Singularity images containing all KERN software; these can be used to deploy all of KERN on any cluster
supporting Singularity\footnote{\url{https://github.com/kernsuite/singularity}}.

\section{Project structure}

All the packaging effort of KERN is strictly open source and open development. All source code is publicly 
available on Github\footnote{\url{http://github.com}}. We use the Github bug tracker to interact with users.
Users can report problems, ask questions and request new features via the bug tracker. Participation is 
encouraged by the means of pull requests. We have a central entry point website which contains a list of common questions
and links to all related services and
pages\footnote{\url{http://kernsuite.info/}}.

\subsection{The release cycle}
We anticipate a KERN release recycle of approximately six months. At the time of writing, the latest release is KERN-2,
released on 13 March 2017. Every KERN release is ``fixed'', in the sense that we do not plan any package updates
post-release,  unless some critical issues need to be addressed. In between KERN releases, development activity proceeds on an
active development branch called KERN-dev. This repository will be constantly updated with new packages, but this should
only be used for testing and experiments rather than for production science.

\subsection{Technical structure}

Our packaging procedure makes extensive use of git. Every release of every package added to KERN is mirrored on our
packaging Github repository\footnote{\url{https://github.com/kernsuite-debian}}. Every new release is a new commit
to the git repository. These commits are then augmented with Debian metadata files. The metadata files contain a description,
build and runtime dependencies as well as a robust script to build, install and clean the package. These metadata files
can be very minimal if properly written build scripts are provided by the original software authors, but can be more
complex in the absence of these.

New packages are published to Launchpad which is a free to use service maintained by Canonical, the company responsible
for the Ubuntu distribution. Packages are built on the Launchpad build farm and we simply upload the base source image.
This is an extra quality check, since the build farm makes sure that the package compiles correctly and that all
the dependencies are correctly defined.

If the original source is provided with (unit)tests, as is the case for casacore, we run these tests
during the creation of the package. Unfortunately, most software packages in KERN do not have a test suite provided
by the developers.

All scripts and packaging files are released under the conditions of the MIT License. The MIT license is a simple
and permissive license, it only requires preservation of copyright and license notices. Licensed works, modifications
and larger works may be distributed under different terms and without source code. Note that this only applies to the
KERN files and not to the software contained by the KERN packages. The licenses of these packages is respected and is
bundled with every package.

\section{Recommended usage}
KERN is freely available and maintained as a service to the community. If you use KERN in a published work, please
state which version of KERN was used and include it as a citation and/or an acknowledgment.

\section{Conclusions}

Although the KERN project does not introduce any novel algorithmic techniques as such, we are of the opinion that
it is a foundational block for a robust radio astronomy software environment. Radio astronomy software has 
historically been difficult to build and install for the end user. Radio astronomers world wide
now use the packages distributed via KERN.

Based on a subjective observation of users within our team at SKA South Africa, ease of software installation was
vastly improved after KERN became available. Subsequently, the KERN ready-to-install packages resulted in a noticeable
reduction in time and effort required to achieve successful software installation.
While we believe this experience will not be unique to SKA SA, a further quantitative survey of the global community
could not be carried out.

We believe that these packages, as well as the regular release cycle,
are a great improvement for reproducible science, problem isolation and discoverability of tools. KERN is not only of
benefit to astronomers, but also to system administrators, who now have to spend less time installing software
and tracking updates and changes.

\paragraph*{Acknowledgments} The research of
O. Smirnov is supported by the South African Research Chairs
Initiative of the Department of Science and Technology and National
Research Foundation.

\section{References}

\bibliography{kern}{} \bibliographystyle{plain}

\newpage

\onecolumn

\appendix

\section{List of packages in KERN}

\begin{longtable}{ l p{0.65\textwidth} r }
\caption{List of packages in KERN}
\label{tab:packagelist} \\

\multicolumn{1}{c}{\textbf{Package name}} &
\multicolumn{1}{c}{\textbf{Description and website}} &
\multicolumn{1}{c}{\textbf{Version in KERN-2}} \\ \hline  \hline
\endfirsthead

\multicolumn{3}{c}%
{{\bfseries \tablename\ \thetable{} -- continued from previous page}} \\
\multicolumn{1}{c}{\textbf{Package name}} &
\multicolumn{1}{c}{\textbf{Description and website}} &
\multicolumn{1}{c}{\textbf{Version in KERN-2}} \\ \hline  \hline
\endhead

\multicolumn{3}{r}{{Continued on next page}} \\
\endfoot

\hline \hline
\endlastfoot

casacore-data &  Casacore data files & \\ & \url{http://casacore.github.io/casacore} & 20171002 \\

\hline
casacore &  Suite of c++ libraries for radio astronomy data processing & \\ & \url{http://casacore.github.io/casacore} & 2.3.0 \\

\hline
21cmfast &  A seminumeric modelling tool to efficiently simulate the cosmological 21-cm signal & \\ & \url{https://github.com/andreimesinger/21cmFAST} & 1.2 \\

\hline
aips &  Calibration, data analysis, image display, plotting, and a variety of ancillary tasks on Astronomical Data & \\ & \url{http://www.aips.nrao.edu/} & 31dec16 \\

\hline
aoflagger &  Find RFI in radio astronomical observations & \\ & \url{http://sourceforge.net/projects/aoflagger/} & 2.9.0 \\

\hline
attrdict &  A dictionary that allows attribute-style access & \\ & \url{https://github.com/bcj/AttrDict} & 2.0.0 \\

\hline
casalite &  Stripped down version of NRAO's CASA & \\ & \url{https://casa.nrao.edu/} & 4.7.1 \\

\hline
casarest &  Standalone radio interferometric imager derived from CASA  & \\ & \url{https://github.com/casacore/casarest/} & 1.4.1 \\

\hline
casasynthesis &  The synthesis CASA 4.4 submodule as a standalone project & \\ & \url{https://github.com/radio-astro/casasynthesis} & 0.1 \\

\hline
cassbeam &  Cassegrain antenna modelling & \\ & \url{https://github.com/ratt-ru/cassbeam} & 1.1 \\
\hline
chgcentre &  Change the phase centre of a measurement set & \\ & \url{https://sourceforge.net/p/wsclean/wiki/chgcentre/} & 1.5 \\
\hline
drive-casa &  A Python package for scripting the NRAO CASA pipeline routines & \\ & \url{https://github.com/timstaley/drive-casa} & 0.7.4 \\
\hline
dspsr &  Library for digital signal processing of pulsar astronomical timeseries & \\ & \url{http://dspsr.sourceforge.net/} & 0+git20170125 \\
\hline
dysco &  A compressing storage manager for Casacore measurement sets & \\ & \url{https://github.com/aroffringa/dysco} & 1.0 \\
\hline
galsim &  Simulating images of astronomical objects (stars, galaxies) in a variety of ways & \\ & \url{https://github.com/GalSim-developers/GalSim} & 1.4.3 \\
\hline
karma &   Toolkit for interprocess communications, authentication, encryption, graphics display, user interface and manipulating the Karma network data structure & \\ & \url{http://www.atnf.csiro.au/computing/software/karma/} & 1.7.25 \\
\hline
katdal &  Data access library for the MeerKAT project  & \\ & \url{https://github.com/ska-sa/katdal} & 0.7.1 \\
\hline
katpoint &  Karoo Array Telescope pointing coordinate library & \\ & \url{https://github.com/ska-sa/katpoint} & 0.6 \\
\hline
katversion &  Provides proper versioning for Python packages & \\ & \url{https://github.com/ska-sa/katversion} & 0.7 \\
\hline
kittens &  Collection of Python utility functions for purr, tigger, meqtrees and others & \\ & \url{https://github.com/ska-sa/kittens} & 1.3.3 \\
\hline
lofar &  LOFAR telescope user software & \\ & \url{http://www.lofar.org} & 2.20.2 \\
\hline
makems &  Tool to create empty Measurement Sets & \\ & \url{https://github.com/ska-sa/makems} & 1.3.0 \\
\hline
meqtrees-cattery &  MeqTrees-based frameworks for simulation and calibration of radio interferometers & \\ & \url{https://github.com/ska-sa/meqtrees-cattery} & 1.5.1 \\
\hline
meqtrees-timba &  Implementing and solving arbitrary Measurement Equations & \\ & \url{http://www.astron.nl/meqwiki/MeqTrees} & 1.5.0 \\
\hline
montblanc &  A PyCUDA implementation of the RIME & \\ & \url{https://github.com/ska-sa/montblanc} & 0.4.0 \\
\hline
msutils &  A set of CASA Measurement Set manipulation tools & \\ & \url{https://github.com/SpheMakh/msutils} & 0.0.2 \\
\hline
mt-imager &  High performance image synthesiser for radio interferometry & \\ & \url{http://www.mtimager.com/} & 1.0 \\
\hline
multinest &  A Bayesian inference tool & \\ & \url{https://github.com/JohannesBuchner/MultiNest} & 2.14+git20150803 \\
\hline
obit &  Obit for ParselTongue & \\ & \url{http://www.cv.nrao.edu/~bcotton/Obit.html} & 22JUN10k \\
\hline
oskar &  Simulator for the Open Square Kilometre Array Radio Telescope & \\ & \url{http://www.oerc.ox.ac.uk/~ska/oskar2/} & 2.6.1 \\
\hline
owlcat &  Miscellaneous utility scripts for manipulating radio interferometry data & \\ & \url{https://github.com/ska-sa/owlcat} & 1.4.2 \\
\hline
parseltongue &  Python scripting interface for classic AIPS & \\ & \url{http://www.jive.nl/jivewiki/doku.php?id=parseltongue:parseltongue} & 2.3 \\
\hline
presto &  A large suite of pulsar search and analysis software & \\ & \url{https://github.com/scottransom/presto/} & 2+git20160604 \\
\hline
psrcat &  The ATNF Pulsar Catalogue & \\ & \url{http://www.atnf.csiro.au/people/pulsar/psrcat/} & 1.55 \\
\hline
psrchive &  C++ development library for the analysis of pulsar astronomical data & \\ & \url{http://psrchive.sourceforge.net/} & 2012.12+20170131 \\
\hline
purify &  Collection of routines for radio interferometric imaging & \\ & \url{http://basp-group.github.io/purify/} & 2.0.0 \\
\hline
purr &  Data reduction logging tool, Useful for remembering reductions  & \\ & \url{https://github.com/ska-sa/purr} & 1.3.1 \\
\hline
pymoresane &  Python version of the MORESANE deconvolution algorithm & \\ & \url{https://github.com/ratt-ru/PyMORESANE} & 0.3.6 \\
\hline
python-casacore &  Python 2 bindings to the casacore library & \\ & \url{https://github.com/casacore/python-casacore} & 2.1.2 \\
\hline
pyxis &  Python Extensions for Interferometry Scripting & \\ & \url{https://github.com/ska-sa/pyxis} & 1.5.1 \\
\hline
rfimasker &  Tool to apply RFI masks & \\ & \url{https://github.com/bennahugo/RFIMasker} & 0.0.1 \\
\hline
rpfits &  Data-recording format & \\ & \url{http://www.atnf.csiro.au/computing/software/rpfits.html} & 2.24 \\
\hline
sagecal &  Very fast, memory efficient and GPU accelerated radio interferometric calibration program & \\ & \url{https://sourceforge.net/projects/sagecal/} & 0.2+20170130 \\
\hline
scatterbrane &  Adding realistic scattering to astronomical images & \\ & \url{https://github.com/krosenfeld/scatterbrane} & 0.0+git20160122 \\
\hline
sigproc &  Pulsar search and analysis software & \\ & \url{https://github.com/SixByNine/sigproc} & 0.0+git20161025 \\
\hline
sigpyproc &  Python-based pulsar search data manipulation package & \\ & \url{https://github.com/ewanbarr/sigpyproc} & 0.0+git20160115 \\
\hline
simfast21 &  Generates a simulation of the cosmological 21cm signal & \\ & \url{https://github.com/mariogrs/Simfast21} & 2.0~beta \\
\hline
simms &  Creates empty measurement sets using the the CASA simulate tool & \\ & \url{https://github.com/radio-astro/simms} & 0.9.1 \\
\hline
sopt &  Sparse OPTimisation shared library & \\ & \url{http://basp-group.github.io/sopt/} & 2.0.0 \\
\hline
sourcery &  Tools for creating high fidelity source catalogues from radio interferometric datasets & \\ & \url{https://github.com/radio-astro/sourcery} & 1.2.6 \\
\hline
spdlog &  Very fast, header only, C++ logging library & \\ & \url{https://github.com/gabime/spdlog} & 1:0.11.0 \\
\hline
stimela &  Dockerized Radio Interferometry Scripting Framework & \\ & \url{https://github.com/SpheMakh/Stimela} & 0.2.7 \\
\hline
tempo &  Pulsar timing data analysis package & \\ & \url{https://sourceforge.net/projects/tempo/} & 0+git20160709 \\
\hline
tempo2 &   Pulsar timing package & \\ & \url{https://bitbucket.org/psrsoft/tempo2/} & 2017.2.1 \\
\hline
tigger &  FITS and MeqTrees LSM viewer & \\ & \url{https://github.com/ska-sa/tigger} & 1.3.7 \\
\hline
tirific &  Simulate kinematical and morphological models & \\ & \url{http://gigjozsa.github.io/tirific/} & 2.3.7 \\
\hline
tkp &  A transients-discovery pipeline for astronomical image-based surveys & \\ & \url{https://github.com/transientskp/tkp} & 4.0 \\
\hline
tmv &  A fast, intuitive linear algebra library for C++ & \\ & \url{https://github.com/rmjarvis/tmv} & 0.74 \\
\hline
transitions & Lightweight, object-oriented finite state machine implementation & \\ & \url{https://github.com/tyarkoni/transitions} & 0.4.3 \\
\hline
wsclean &  Fast generic widefield interferometric imager & \\ & \url{http://sourceforge.net/projects/wsclean/} & 2.3 \\

\end{longtable}

\end{document}